\documentclass[twocolumn,twoside,slac_two]{revtex4}
\usepackage{graphicx}
\usepackage{fancyhdr}
\begin{document}

\title{The Fermi Motion and $J/\psi$ Production at
the Hadron Colliders }

%

\author{M.A. Gomshi Nobary, B. Nikoobakht}
\affiliation{Department of Physics, Faculty of Science, Razi
University, Kermanshah, Iran.}
\begin{abstract}
The Fermi motion effect in $J/\psi$ production in various hadron
colliders is studied. We deduce that in agreement with sum rules
which the fragmentation functions should satisfy, while the effect
has considerable influence on the fragmentation probabilities and
the differential cross sections, the total cross sections are
essentially left unchanged.
\end{abstract}

\maketitle

\thispagestyle{fancy}

\section {INTRODUCTION}
The $J/\psi$ state has been one of the interesting problems of QCD
both in theory [1,2,3] and in experiment [4]. QCD predictions for
production cross section of $J/\psi$ fails to agree with the
experimental results. Color octet senario is introduced to bring
about the agreement [5,6].

In this work, we introduce the Fermi motion into the $J/\psi$
production. We only include the charm fragmentation contributions.
The Tevatron Run I, the Tevatron Run II, the RHIC and the CERN LHC
energies are considered to demonstrate the effect.

\section {FRAGMENTATION FUNCTIONS}

With the choice of a light cone the wave function, we introduce
the Fermi motion effect in fragmentation production of the
$J/\psi$ bound state. In the leading order perturbative regime,
the fragmentation functions for $J/\psi$ production without and
with the Fermi motion are obtained [7].

In the absence of Fermi motion, where the confinement parameter is
$\beta=0$, the fragmentation function has the following form

\begin{widetext}
\begin{eqnarray}
D_{c\rightarrow J/\psi}(z,\mu_\circ,\beta=0)& =&{{\alpha_s^2
C_F^2\langle {k_T}^2\rangle^{1/2}}\over 16 m { F(z)}}\biggl\{
z(1-z)^2\Bigl[\xi^2z^4+2\xi z^2(4-4z+5z^2)\nonumber\\
&&+(16-32z+24z^2-8z^3+9z^4)\Bigr]\biggr\},
\end{eqnarray}
\end{widetext}
where $\alpha_s$ is the strong interaction coupling constant and
$C_F$ is the color factor. The quantity $ \langle {k_T}^2\rangle$
is the average transverse momentum squared of the initial state
heavy quark, the parameter $\xi$ is defined as $ \xi=\langle
{k_T}^2\rangle/m^2 $ and finally $F(x)$ is given by

\begin{widetext}
\begin{eqnarray}
{ F(z}) =\Bigl[\xi^2z^4-(z-2)^2(3z-4)+\xi z^2(8-7z+z^2)\Bigr]^2.
\end{eqnarray}
\end{widetext}

In the presence of Fermi motion the fragmentation function is
obtained as
\begin{widetext}
\begin{eqnarray}
D_{c\rightarrow J/\psi}(z,\mu_\circ,\beta)&
=&{{\pi^2\alpha_s^2C_F^2\langle {k_T}^2\rangle^{1/2}}\over {2 m
}}\int \frac{dq dx
|\psi_M|^2x^2(1-z)^2 zq}{{ G(z})}\nonumber\\
&&\times\Bigl\{1-4(1-x)z+2(4-10x+7x^2)z^2\nonumber\\
&&+4(-1+x^3-5x^2+4x)z^3+(1-4x+8x^2-4x^3+x^4)z^4\nonumber\\
&&+\eta\xi z^2\bigl[1-2x+z^2+x^2(2-2z+z^2)\bigr]+ \eta\bigl[
2+(-6+4x)z\nonumber\\
&&+(9-8x+2x^2)z^2-2(2-x+x^2)z^3+(1+x^2)z^4\bigr]\nonumber\\
&&+\xi z^2\bigl[1+2x^3(2-3z)z+z^2+2x^4z^2+2x(-1+z-2z^2)\nonumber\\
&& +x^2(2-8z+9z^2)\bigr]+\eta^2(1-z)^2+\xi^2(1-x)^2x^2z^4\Bigr\}.
\end{eqnarray}
\end{widetext}
The function ${ G(z)}$ reads as

\begin{widetext}
\begin{eqnarray}
{ G(z)}&=&\Bigl\{\bigl[\eta (1-z)^2+\xi
x^2z^2+(1-(1-x)z)^2\bigr]\bigl[\eta(-1+z)+\xi
(-1+x)xz^2-1+(1-x+x^2)z\bigr]\Bigr\}^2.
\end{eqnarray}
\end{widetext}
Moreover here we have defined $\eta=q^2/m^2$.

The two cases of the Fermi motion off and on may be compared using
the above fragmentation functions. Note that $\beta=0$ and
$\beta\neq 0$ correspond to these cases.  For  $\beta$ value, we
have chosen the range of $\beta=$0.0 - 0.6 GeV . The behavior of
the fragmentation function (3) is shown in Fig. 1 for $\beta=0,\;
0.2, \;0.4 \;{\rm and} \;0.6$ GeV. We take the maximum value of
$\beta$=0.6 GeV and use it for our further considerations.

\begin{figure}
\begin{center}
\includegraphics[width=8.5cm]{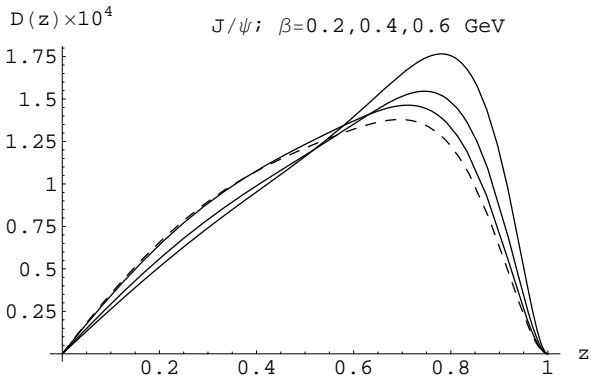}
\caption{Fragmentation function for $J/\psi$ production. While the
dashed curve represents the function when the Fermi motion is off,
the solid ones show the behavior of this function when the Fermi
motion is on. The $\beta$ values are indicated. The function picks
up as $\beta$ increases. This gives raise to increase in the
fragmentation probability. }
\end{center}
\end{figure}

\section {INCLUSIVE PRODUCTION CROSS SECTION }

We have employed the idea of factorization to evaluate the
$J/\psi$ production cross section at hadron colliders. For $\bar
{p} p$ collisions we may write

\begin{widetext}
\begin{eqnarray}
\frac {d\sigma}{dp_T}(\bar{p} p \rightarrow c \rightarrow J/\psi
(p_T) X)&=&\sum_{i,j}\int dx_1 dx_2 dz f_{i/\bar{p}}(x_1,\mu)f_{j/
p}(x_2,\mu)\nonumber\\&&\times\Bigl[ \hat\sigma(ij\rightarrow
c(p_T/z)X,\mu) D_{ c\rightarrow J/\psi}(z,\mu,\beta)\Bigr].
\end{eqnarray}
\end{widetext}

Where $f_{i,j}$ are parton distribution functions with momentum
fractions of $x_1$ and $x_2$, $\hat\sigma$ is the charm quark hard
production cross section and $D(z,\mu,\beta)$ represents the
fragmentation of the produced heavy quark into $\bar c c$ state
with confinement parameter $\beta$ at the scale $\mu$. We have
used the parameterization due to Martin-Roberts-Stiriling (MRS)
[8] for parton distribution functions and have included the heavy
quark production cross section up to the order of $\alpha_s ^3$
[9]. The dependence on $\mu$ is estimated by choosing the
transverse mass of the heavy quark as our central choice of scale
defined by
\begin{eqnarray}
\mu_R=\sqrt{ {p_T}^2{\rm (parton)}+{m_c}^2,   }
\end{eqnarray}
and vary it appropriate to the fragmentation scale of the particle
state to be considered. This choice of scale, which is of the
order of $p_T$ (parton), avoids the large logarithms in the
process of the form $\ln(m_Q/\mu)$ or $\ln(p_T/\mu)$. However, we
have to sum up the logarithms of order of $\mu_R/m_Q$ in the
fragmentation functions. But this can be implemented by evolving
them by the Altarelli-Parisi equation [10]. The following form of
this equation is used here

\begin{widetext}
\begin{eqnarray}
\mu {\partial\over{\partial \mu}} D_{c\rightarrow J/\psi}
(z,\mu,\beta)= \int_z^1 {dy\over y} P_{Q \rightarrow Q} (z/y,\mu)
D_{c\rightarrow J/\psi}(y,\mu,\beta).
\end{eqnarray}
\end{widetext}

 Here $P_{Q \rightarrow Q}(x=z/y,\mu) $ is the Altarelli-Parisi
splitting function. The boundary condition on the evolution
equation (7) is the initial fragmentation function
$D_{c\rightarrow J/\psi}(z,\mu,\beta)$ at some scale
$\mu=\mu_\circ$. In principle this function may be calculated
perturbatively as a series in $\alpha_s$ at this scale.

Detection of final state requires kinematical cuts of the
transverse momentum, $p_T$, and the rapidity, $y$. We have imposed
the required $p_T^{\rm cut}$ and $y^{\rm cut}$ in our simulations
for different colliders as required and have used the following
definition of rapidity $y=\frac{1}{2} \log[(E-p_L)/(E+p_L)]$.

\begin{figure}
\begin{center}
\includegraphics[width=8.5cm]{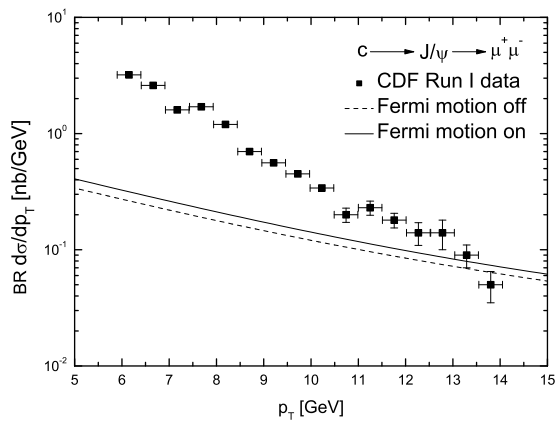}
\caption{The differential cross section for direct fragmentation
production of $J/\psi$ and its subsequent decay $J/\psi
\rightarrow \mu^+ \mu^-$ at the Tevatron Run I energies. While the
dashed curve is obtained using (1) or equally (3) with $\beta=0$,
the solid one is due to (3) with $\beta=0.6$ GeV. The result is
compared with the CDF Run I data. Other contributions are not
included. The scale is chosen to be $2\mu_R$. }
\end{center}
\end{figure}

\begin{figure}
\begin{center}
\includegraphics[width=8.5cm]{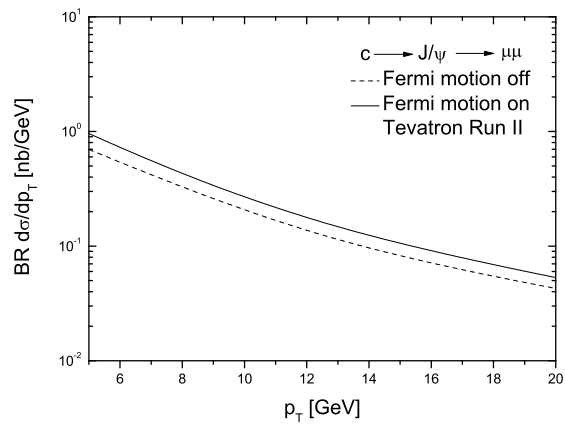}
\caption{The differential cross section for direct fragmentation
production of $J/\psi$ at Tevatron Run II. The two curves are
obtained using (3) with $\beta= 0$ and 0.6 GeV respectively. The
scale has been set to $2\mu_R$. }
\end{center}
\end{figure}

\begin{figure}
\begin{center}
\includegraphics[width=8.5cm]{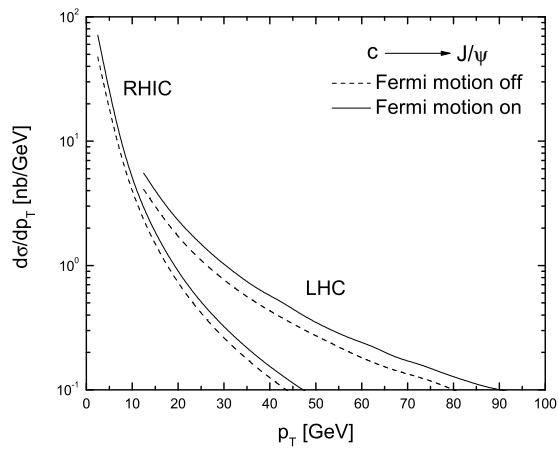}
\caption{The differential cross section for direct fragmentation
production of $J/\psi$ at the RHIC and the CERN LHC. The shift due
to the Fermi motion is significantly increased in the case of LHC
but shows to be less important at the RHIC. In both cases the
fragmentation function (3) is employed with $\beta= 0$ and 0.6 GeV
respectively.}
\end{center}
\end{figure}

\section {RESULTS AND DISCUSSION}

In this work we have employed a light-cone wave function to
introduce the Fermi motion in $J/\psi$ production in direct
fragmentation channel and obtained its fragmentation function in
leading order perturbative regime. The behavior of $J/\psi$
fragmentation function with such a wave function is illustrated in
Fig. 1.

We have studied the production rates at which the $J/\psi$ state
is produced with and without the effect of Fermi motion. First we
present the $p_T$ distribution of ${\rm BR}
(J/\psi\rightarrow\mu^+\mu^-)d\sigma/dp_T$ for the cases of the
Fermi motion off and on along with the CDF data at Run I in the
Fig. 2. The branching ratio BR($J/\psi\rightarrow
\mu^+\mu^-$)=0.0597 is taken from [11]. The poor agreement with
data is due to the fact that here we have only considered the
contribution of $\bar{p}p\rightarrow c\rightarrow J/\psi
\rightarrow \mu^+ \mu^-$. Similar behavior at Tevatron Run II
energies is shown in Fig. 3. We have also extended our study to
the cases of the RHIC and the CERN LHC $ p p$ colliders. Here we
provide the $p_T$ distributions of the differential cross sections
for $c\rightarrow J/\psi$ and compare the two cases of the Fermi
motion off and on in the Fig. 4. In all cases we have used
$\beta=0.6$ GeV for the confinement parameter in the fragmentation
functions. Naturally, the results for $\beta$ in the range of 0 -
0.6 GeV fall between the above results.

Calculation of the total integrated cross sections for each case
shows that the total cross sections for with and without Fermi
motion essentially remain unchanged within the uncertainties of
Mote Carlo simulations. The reason is first due to the momentum
sum rule which the fragmentation functions should satisfy. In
other words although the modification of fragmentation functions
by the Fermi motion redistributes the final states, the integrated
cross sections are left unchanged. Alternatively although the
Fermi motion increases the fragmentation probability for the
state, i.e., introduces a state with overall higher mass, the
cross section is lowered by just the same amount when we introduce
the effect in calculation of the total integrated cross section.
It is evident from the Figures 2,3 and 4 that the effect increases
with increasing $\sqrt{s}$. The kinematical cuts play important
role apart from $\sqrt {s}$. The large cross section at the RHIC
compared with the LHC in the Fig. 4 is an example.

There are two main sources of uncertainities.The first is about
the simulation of $J/\psi$ production at hadron colliders such as
the uncertainties along with the fragmentation functions and
parton distribution functions. These kind of uncertainties are
well discussed in the literature. The second source of uncertainty
is due to the choice of the confinement parameter in the
fragmentation function. Relying on our discussion in section 2,
our choice of $\beta=0.6$ GeV seems to be justified. Future
determination of this parameter will shed more light on this
situation. It is worth mentioning that our choice of charm quark
mass, i.e., 1.25 GeV, have put our results in their upper side and
that the change of the charm quark mass  in its acceptable range
does not have significant impact on the Fermi motion effect in
$J/\psi $ production.

More detailed information concerning this work appears in [12].

\end{document}